\documentclass[conference]{IEEEtran}
\IEEEoverridecommandlockouts
\usepackage{cite}
\usepackage{amsmath,amssymb,amsfonts}
\usepackage{algorithmic}
\usepackage{graphicx}
\usepackage{textcomp}
\usepackage{titlesec}
\usepackage{xcolor}
\def\BibTeX{{\rm B\kern-.05em{\sc i\kern-.025em b}\kern-.08em
    T\kern-.1667em\lower.7ex\hbox{E}\kern-.125emX}}

\usepackage[all]{background}
\usepackage{stackengine}
\setstackEOL{\\}
\setstackgap{L}{\normalbaselineskip}
\SetBgContents{\color{gray}{\tiny \Longstack{PREPRINT - Accepted as PhD Forum at VLSI-SoC 2024}}}
\SetBgPosition{4.5cm,1cm}
\SetBgOpacity{1.0}
\SetBgAngle{0}
\SetBgScale{1.8}

\begin{document}

\setlength\abovecaptionskip{0\baselineskip}
\setlength\belowcaptionskip{-1.4\baselineskip}
 
\def\@IEEEfigurIEEEecaptionsepspace{\vskip\abovecaptionskip\relax}%
\def\@IEEEtablecaptionsepspace{\vskip\abovecaptionskip\relax}%

\titlespacing*{\section}
{0pt}{1pt}{0pt}

\title{Resistive Memory for Computing and Security: Algorithms, Architectures, and Platforms\vspace{-0.5cm}}

\author{Simranjeet Singh\IEEEauthorrefmark{1}, Farhad Merchant\IEEEauthorrefmark{2}, Sachin Patkar\IEEEauthorrefmark{1} \\
\IEEEauthorrefmark{1}IIT Bombay, India,\IEEEauthorrefmark{2}Newcastle University, UK \\ 
\{simranjeet, patkar\}@ee.iitb.ac.in, farhad.merchant@newcastle.ac.uk
\vspace{-0.5cm}}

\maketitle

\begin{abstract}
Resistive random-access memory (RRAM) is gaining popularity due to its ability to offer computing within the memory and its non-volatile nature. The unique properties of RRAM, such as binary switching, multi-state switching, and device variations, can be leveraged to design novel techniques and algorithms. This thesis proposes a technique for utilizing RRAM devices in three major directions: i) digital logic implementation, ii) multi-valued computing, and iii) hardware security primitive design. We proposed new algorithms and architectures and conducted \textit{experimental studies} on each implementation.  Moreover, we developed the electronic design automation framework and hardware platforms to facilitate these experiments.

\end{abstract}

\begin{IEEEkeywords}
RRAM, digital logic-in-memory, multi-valued logic, hardware security
\end{IEEEkeywords}

\section{Motivation}
The growing popularity of resistive random-access memory (RRAM) is fueled by its remarkable capability to store multi-bit and embed computing functionalities directly within the memory. RRAM's distinctive properties, particularly its ability to exhibit multi-state behavior and stochasticity, open up new avenues for innovation in the realm of algorithm, architecture, and security design. RRAMs generally consist of a transition metal oxide (TMO) layer sandwiched between top and bottom electrodes in a metal-insulator-metal configuration. The resistance of the TMO layer can modulated using external electrical signals, facilitating data storage in distinct resistive states. Typically, RRAM devices operate in a binary switching mode, where they exhibit two distinct states: a low-resistance state (LRS) representing logic `1’ and a high-resistance state (HRS) representing logic `0’~\cite{bende2024experimental}. However, the multi-state mode can be achieved by inducing a gradual change in the RESET voltage, leading to a progressive increase in resistance between LRS and HRS rather than an abrupt transition. It is important to note that RRAM devices have a significant impact on the resistance state of the device due to the device-to-device (D2D) and cycle-to-cycle (C2C) variations~\cite{singh2023hs}. 




Taking into account the Boolean, multi-state switching, and stochastic nature of RRAM, this thesis endeavors to harness the full potential of RRAM by introducing novel methodologies for its utilization. The thesis emphasizes i) digital logic-in-memory (LiM), ii) multi-valued logic (MVL) computing, and iii) the design of hardware security primitives. Additionally, we created an electronic design automation (EDA) framework and hardware prototype centered on RRAM to facilitate experimentation. The thesis explores RRAM's capabilities in these domains and showcases its adaptability and efficiency in tackling modern computing challenges.

\begin{figure}[t]
    \centering
    \includegraphics[width=\linewidth]{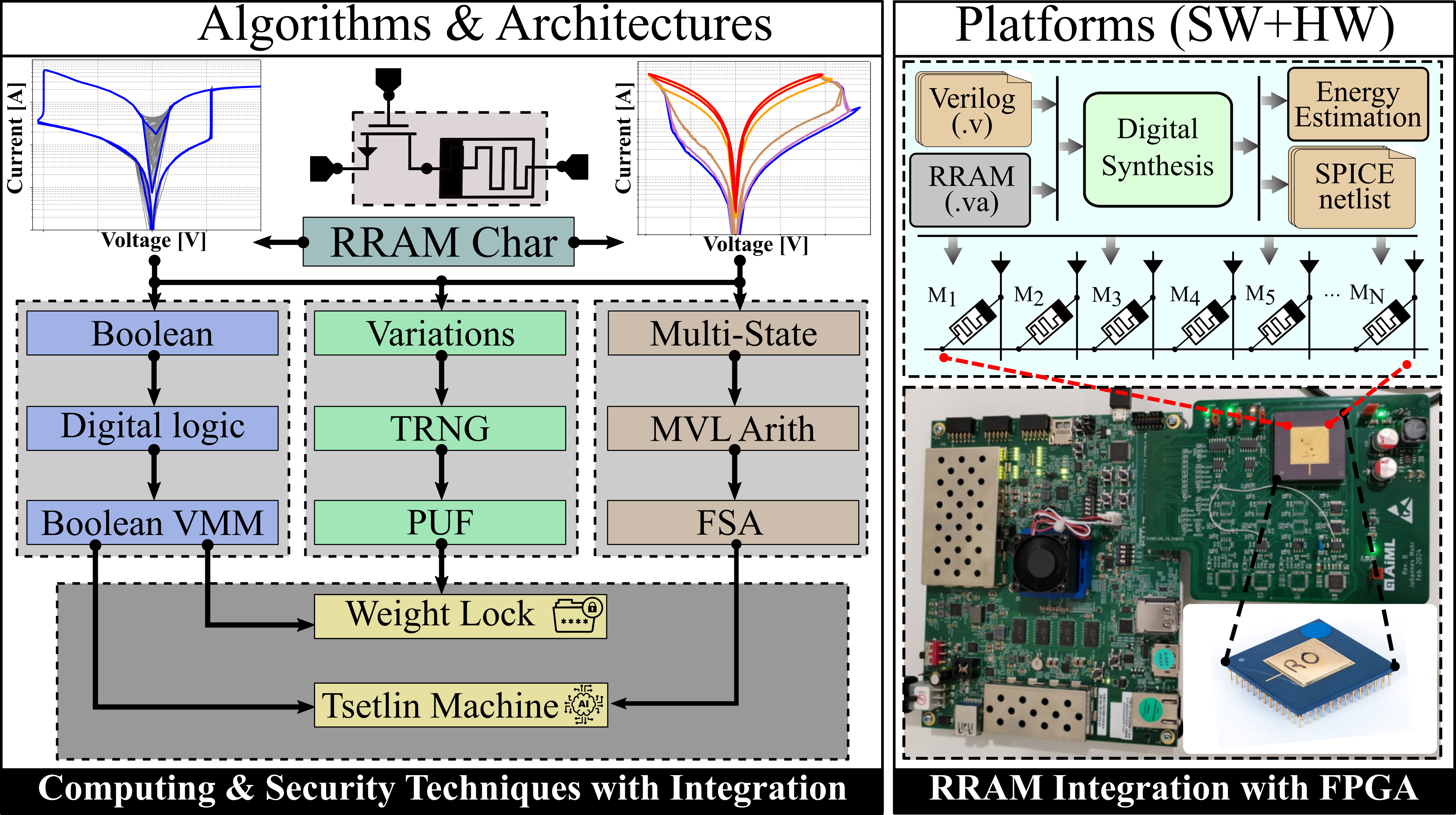}
    \caption{RRAM for computing and security\vspace{-0.5cm}}
    \label{fig:outline}
\end{figure}

\textbf{Contributions:} The contributions of the underlying thesis are as follows (visualized in Fig.~\ref{fig:outline}): 

\begin{itemize}
\item Designing novel logic gates using the Boolean properties of RRAM and conducting comprehensive experimental studies to validate its efficacy.
\item Designing multi-level arithmetic and finite state automata (FSA) utilizing multi-state properties and developing architectures for the Tsetlin machine.
\item Designing architectures and algorithms for i) implementing true random number generators (TRNG) and physical unclonable functions (PUF) on a single RRAM crossbar and ii) introducing a technique for securely locking neural network weights utilizing an integrated hardware security module.
\item Creating an EDA framework for synthesizing hardware description language (HDL) into SPICE-level netlists for accurate energy analysis.
\item Integrating packaged RRAM chips with FPGAs for hardware prototyping.
\end{itemize}



\section{Digital LiM}
Considering Boolean properties, we initially implemented logic gates using the VTEAM model and evaluated their energy consumption. The analysis revealed the prominence of initialization energy in digital LiM~\cite{singh2023ESL,jha2024verification}. This prompted the exploration of novel logic gate designs on a different material stack with distinct properties from VTEAM, such as HRS/LRS and SET/RESET ratio. Devices fabricated with the TaOx switching stack possess these properties and are supported by an experimentally validated model (JART VCM). Utilizing this model, we proposed novel in-memory cloning methods~\cite{singh2024inmemorymirroringcloningreading} and logic gate designs, including OR and NOT gates, on the TaOx 1T1R crossbar. Subsequently, we experimentally validated the proposed gates on a fabricated 8x4 TaOx 1T1R crossbar and observed energy consumption trends~\cite{bende2024experimental}. Next, we moved to the multi-valued logic design, considering the multi-state properties of RRAM. 


\section{MVL Computing}
As a first step, we investigated state-switching dynamics to facilitate MVL operations. Employing a gradual RESET method on TaOx devices, we achieved multi-level behavior, resulting in at least six stable states between the LRS and HRS. These states were leveraged to develop a ternary arithmetic adder, demonstrated by the 41-trit adder (equivalent to a 64-bit digital adder)~\cite{singh2023nanorarch}. In addition to arithmetic circuits, we proposed the design of FSA utilizing the MVL properties, alongside investigating the impact of D2D and C2C variations on state transitions and detection. The FSA implementation is then simulated, showcasing the Krinsky learning automaton~\cite{singh2023FSA}. 

Finally, we introduced a Tsetlin machine inference architecture that harnesses the Boolean and multi-state properties of RRAM altogether~\cite{ghazal2023imbue}. The proposed architecture demonstrated significant accuracy and energy efficiency improvements compared to traditional machine learning implementations.


\section{Hardware Security}
Variations such as D2D and C2C pose challenges in computing algorithms, as they significantly affect resistance states, leading to computing errors. While schemes like checksum aim to mitigate errors during in-memory computing due to variation~\cite{parrini2024error}. However, we leverage these variations to design a hardware security module. We propose TRNG and PUF architecture on the same RRAM crossbar, where the TRNG generates entropy utilized by the PUF~\cite{singh2023hs}. Subsequently, we propose multiple configurations of these designs to enhance PUF metrics~\cite{singh2022PAPUF,ranjendra2024PRPUF,ranjendra2024harnessing}. Finally, we integrate the PUF with a neural network (NN) on the same crossbar to secure the non-volatile NN weights~\cite{singh2023NNlock}. 


\section{EDA and Hardware Prototype}
Next, we developed an EDA framework that considered the different aspects of implementation and many existing RRAM models. For digital LiM, The framework allows the synthesis of the HDL to LiM design at the SPICE level. The proposed framework automatically generates a SPICE-level netlist and testbench voltages for a given application/benchmark. Furthermore, it provides fine-grained energy numbers by calculating the energy consumed by each device in the crossbar~\cite{singh2024memspice}. The framework empowers researchers to obtain accurate energy estimates for digital designs, offering valuable insights into their methodologies at the circuit level. The framework also performs the D2D and C2C variation simulation for hardware security, especially for PUF and TRNG analysis. 

We developed a hardware prototype by integrating the RRAM packaged chip fabricated by CEA-Leti, France, with the FPGA, as shown in Fig.~\ref{fig:outline}. The packaged chip has multiple crossbar configurations with digital selection peripherals, the biggest crossbar size being $512 \times 32$. We integrated the analog peripherals, such as sense amplifiers, ADCs, DACs, and analog switches outside the chip, to process the signals for any given application. The integration supports custom instructions that a designed FPGA controller handles. The hardware prototype was demonstrated at embedded word 2024~\cite{embeddedword}.



\section{Conclusion \& research directions}
In conclusion, this thesis leverages the RRAM properties to design novel algorithms and architectures for computing and security. The proposed concepts are experimentally validated, providing the EDA and hardware prototype tools to facilitate the experimental studies across domains. 

Expanding upon the insights gained from our research, our goal is to investigate a hybrid approach capable of integrating various schemes into a reconfigurable crossbar. This crossbar concept consolidates all peripherals and functionalities for Boolean, multi-state, and security operations within a single crossbar, facilitating the hybrid approach to cater to a wide range of applications. Also, we aim to enhance the hardware prototype by incorporating all peripherals onto the packaged chip and EDA for hardware prototyping. This initiative lays the foundation for developing next-generation architectures leveraging RRAM technology.

\section*{Acknowledgment}
I sincerely thank Mr. Johannes Mohr for his invaluable contribution to revising the hardware platform. Additionally, I am deeply thankful to Dr. Vikas Rana for inviting me to the Forschungszentrum Jülich, Germany, for a year-long research~visit.

\bibliographystyle{IEEEtran}

\end{document}